\documentclass[aps,prd,twocolumn,superscriptaddress]{revtex4}

\usepackage{graphics}

\def\LL{\left\langle}   
\def\RR{\right\rangle}  
\def\LP{\left(}         
\def\RP{\right)}        
\def\LS{\left[}         
\def\RS{\right]}        

\def\BE{\begin{equation}}
\def\EE{\end{equation}}
\def\BEA{\begin{eqnarray}}
\def\EEA{\end{eqnarray}}
\def\EL{\nonumber\\}

\begin{document}


\title{A lattice calculation of vector meson couplings to the vector and tensor currents using chirally improved fermions}


\author{ V. M. Braun }
\author{ T. Burch }
\author{ C. Gattringer }
\affiliation{Institut f\"ur Theoretische Physik, Universit\"at Regensburg,
D-93040 Regensburg, Germany}

\author{ M. G\"ockeler }
\affiliation{Institut f\"ur Theoretische Physik, Universit\"at Leipzig,
D-04109 Leipzig, Germany}
\affiliation{Institut f\"ur Theoretische Physik, Universit\"at Regensburg,
D-93040 Regensburg, Germany}

\author{ G. Lacagnina }
\author{ S. Schaefer }
\author{ A. Sch\"afer }
\affiliation{Institut f\"ur Theoretische Physik, Universit\"at Regensburg,
D-93040 Regensburg, Germany}

\collaboration{for the BGR [Bern-Graz-Regensburg] Collaboration}

\date{June 6, 2003}

\begin{abstract}
We present a quenched lattice calculation of $f_V^\perp/f_V$, the coupling 
of vector mesons to the tensor current normalized by the vector meson decay 
constant. The chirally improved lattice Dirac operator, which allows us 
to reach small quark masses, is used. We put emphasis on analyzing the 
quark mass dependence of $f_V^\perp/f_V$ and find only a rather weak 
dependence. Our results at the $\rho$ and $\phi$ masses agree well with 
QCD sum rule calculations and those from previous lattice studies. 
\end{abstract}


\maketitle

\section{Introduction}

It is widely accepted that the rich spin structure of hard exclusive
processes involving light vector $\rho, K^*,\phi$ mesons provides a 
number of nontrivial possibilities to study the underlying 
short-distance dynamics. One example is given by 
vector meson electroproduction at large virtualities or large momentum 
transfers at HERA \cite{Crittenden:2001tn,Ciborowski:2002qx,Chekanov:2002rm}, 
HERMES \cite{Hartig:2000yw}, and, in the future, COMPASS. In these cases the 
theoretical predictions for 
the production of longitudinally and transversely polarized
mesons are very different \cite{Brodsky:1994kf}. 
Other examples are provided 
by exclusive semileptonic $B\to \rho \ell \nu_\ell$, rare radiative
$B\to\rho\gamma$ or nonleptonic, $B\to \pi\rho$ etc. decays of $B$-mesons,
which are attracting continuous interest as the prime source of
information about the CKM mixing matrix; see, e.g., 
Ref.~\cite{Battaglia:2003in} 
for an exposition of recent developments. In all cases, disentangling the 
longitudinally and transversely polarized (final) vector-meson states 
proves to be crucial since these cases often correspond to 
different underlying weak interaction physics. The theoretical 
description of such processes 
is developing rapidly and thus requires a more accurate and 
reliable determination of the relevant nonperturbative parameters.

From the vector meson side, the QCD description involves vector meson 
distribution amplitudes \cite{Efremov:1980qk,Lepage:1980fj,Chernyak:1984ej} 
which
correspond to probability amplitudes 
for finding the quark and the antiquark in the meson with given momentum
fractions and a small transverse separation. The distribution amplitudes
for longitudinally polarized and transversely polarized vector mesons 
are different and their normalization (i.e. the integral over the
momentum fractions) is given by the matrix elements of the vector
and the tensor current
\BEA
\label{VVeq}
\LL 0 \left| \bar q \gamma^\mu q \right| V(p;\lambda) \RR &=& 
m_V^{} \, f_V^{} \, e_\lambda^\mu \; ,
\\
\label{VTeq}
\LL 0 \left| \bar q \sigma^{\alpha\beta} q \right| V(p;\lambda) \RR &=& 
i \, f_V^\perp(\mu) 
\LP e_\lambda^\alpha p^\beta - e_\lambda^\beta p^\alpha \RP \; ,
\EEA
respectively. Here $V(p;\lambda)$ is a generic light vector meson with 
momentum $p$ and polarization vector $e_\lambda^\alpha$ 
such that $e_\lambda^\alpha p_\alpha =0$, $p^2 = m_V^2$. 
Also $\sigma^{\alpha\beta}= \frac{i}{2}[\gamma^\alpha,\gamma^\beta]$ 
and $q$ is the light quark field,
$q=u,d,s \ldots \;$. In Eqs.~(\ref{VVeq}), (\ref{VTeq}) we suppress 
the isospin structure, for brevity. 
The vector couplings $f_V$ are known from the experimental measurements
in leptonic decays \cite{Hagiwara:2002pw}, while the tensor couplings 
$f_V^\perp$ have to be calculated using some nonperturbative
approach. In particular, from QCD sum rules \cite{Ball:1996tb}
one obtains
\BE
  f_\rho^\perp \; = \; 160\pm 10~\mbox{\rm MeV} \, ,
\label{BBfrho}
\EE
($157 \pm 5$ MeV in Ref.~\cite{Bakulev:1999gf})
and in the applications usually a very weak dependence on the quark mass of 
the ratios was assumed \cite{Chernyak:1984ej,Ball:1998kk}
\BE
   \frac{f_\rho^\perp}{f_\rho} \; \simeq \; \frac{f_{K^*}^\perp}{f_{K^*}}
   \; \simeq \; \frac{f_\phi^\perp}{f_\phi} \;\,.
\label{BBratio}
\EE
(An earlier QCD sum rule determination of $f_\rho^\perp$ in 
Refs.~\cite{Chernyak:1983it,Chernyak:1984ej} suffers from a sign error, 
see Ref.~\cite{Ball:1996tb}.)
At this place it is necessary to add that the couplings $f_V^\perp$, in
contrast to $f_V$, are scale dependent. The corresponding
anomalous dimension is equal to \cite{Shifman:1981dk,Kumano:1997qp,Hayashigaki:1997dn}
\BEA
  \gamma^T &=& \frac{\alpha_s}{2\pi}  C_F +  
              \left(\frac{\alpha_s}{2\pi}\right)^2 C_F
              \left(\frac{257}{36} C_A -\frac{19}{4}C_F-\frac{13}{18}N_f\right)
              \EL && + \; {\cal O}(\alpha_s^3) + ... \; .
 \label{anom-dim}
\EEA
(The three-loop terms \cite{Gracey:2000am}, left out above simply for brevity,
are included in our analysis.)
The number in (\ref{BBfrho}) is given for a low normalization point of 
$\mu=1$~GeV.

The error given in (\ref{BBfrho}) does not include 
intrinsic uncertainties of the sum rule method itself which are difficult to quantify.
Therefore, in view of the importance of this nonperturbative input for phenomenology,
an independent confirmation of these numbers in a lattice calculation with controllable
errors is extremely welcome. Earlier attempts were presented in 
Refs.~\cite{Capitani:1999zd,Becirevic:2003pn} and the results agree well 
with the numbers from sum rules 
(there is also a current effort by the QCDSF 
Collaboration to determine $f_\rho^\perp$; the preliminary result also agrees 
with the sum rule result \cite{QCDSF:2003zz}). 
These calculations were done with ${\cal O}(a)$ improved Wilson fermions. 
With this choice, however, one is restricted to relatively heavy 
pseudoscalars and the smallest
pseudoscalar-mass to vector-mass ratio reached in \cite{Becirevic:2003pn} 
is $m_{PS}/m_V$ = 0.56. 

Here we present a new calculation of the matrix elements which is complementary
to the first lattice results \cite{Becirevic:2003pn}. We use the recently
developed Chirally Improved (CI) Dirac operator \cite{Gattringer:2000js,
Gattringer:2000qu}. It is based on the Ginsparg-Wilson relation 
\cite{Ginsparg:1982bj} and has been shown 
\cite{Gattringer:2002xg,Gattringer:2002sb,Gattringer:2003bp}
to allow simulations with pseudoscalar-mass to vector-mass ratios down to 
$m_{PS}/m_V = 0.28$ at relatively small cost.

In the results we present here we include lattice data down to 
$m_{PS}/m_V$ = 0.33
and thus push our calculation considerably closer to the chiral limit. 
Also at the heavy end we have additional data points such that our quark
masses cover a range of $m_{PS}/m_V$ more than twice as large as the range in 
\cite{Becirevic:2003pn}. This allows us to check the weak quark-mass dependence
of $f_V^\perp/f_V$ predicted by QCD sum rules. The data are compatible
with a linear behavior in the quark mass, respectively in $m_{PS}^2$, and the
slope is small. 
Only for our lightest quarks we find a deviation, which is, however, a
finite size effect. Our final numbers 
agree well both with the QCD sum rule results and the lattice calculations 
in Refs.~\cite{Capitani:1999zd,Becirevic:2003pn}.

\section{Parameters of the calculation}

We perform a quenched calculation with configurations from 
the L\"uscher-Weisz gauge action \cite{Luscher:1985xn,Curci:1983an}
and one step of HYP blocking \cite{Hasenfratz:2001hp}. The final numbers we
quote were computed on $16^3 \times 32$ lattices at two different values of
the gauge coupling, giving rise to lattice spacings of $a = 0.15$ and $a=0.10$
fm \cite{Gattringer:2001jf}. For these lattices the parameters are listed in 
Table~\ref{sim_details}. In addition we performed a series of tests on smaller 
lattices ($12^3\times24$, $8^3\times24$) and at larger cutoff ($a = 0.08$ fm).
This serves to analyze the influence of finite volume and 
to study the scaling behavior.

We compute fermion propagators using the CI Dirac operator at 11 different 
bare quark mass values ranging from $am = 0.01$ to $am = 0.20$.
These quark masses cover a range of $m_{PS}/m_V = 0.33$ to $m_{PS}/m_V = 0.92$.
We do not encounter exceptional configurations and could work at even smaller 
quark masses. However, as we will show, decreasing the quark mass further, thus
going closer to the chiral limit, is not sensible for a calculation of the 
matrix elements at the current physical volumes due to finite size effects. 

In \cite{Gattringer:2002xg,Gattringer:2002sb,Gattringer:2003bp} 
it is demonstrated that the CI operator is effectively 
${\cal O}(a)$ improved. Furthermore the spectroscopy results presented there
show only a very small variation in $a^2$ and thus are essentially free of 
scale dependence. However, improved operators are not yet available for the
CI operator and thus for the matrix elements computed here one can expect only 
linear scaling in $a$ for our final results.   

\begin{table}[b]
\begin{ruledtabular}
\begin{tabular}{cccccc}
$\beta$ & $a$ (fm) \cite{Gattringer:2001jf} & $L$ (fm) & \# confs. 
& $am_q$ & $m_{PS}^{}/m_V^{}$ \cite{Gattringer:2002sb} \\ \hline
7.90 & 0.15 & 2.4 & 100 & 0.02 - 0.20 & 0.38 - 0.85 \\
8.35 & 0.10 & 1.6 & 100 & 0.01 - 0.20 & 0.33 - 0.92 \\
\end{tabular}
\end{ruledtabular}
\caption{
\label{sim_details}
Parameters for our large lattices ($16^3 \times 32$).
}
\end{table}

The renormalization constants necessary for our observables
were calculated according to the
method presented in \cite{Martinelli:1995ty,Gockeler:1998ye}. The procedure 
is patterned after the definition used in the continuum. In a fixed gauge 
(we use the Landau gauge)
the numerically evaluated amputated Green's functions are compared to their 
tree level counterparts and the renormalization constants are read off.
The resulting numbers are in the so-called RI-MOM scheme and are converted 
to $\overline{\mbox{MS}}$ using the perturbative matching coefficients. 
For the vector and tensor renormalization constants, however, these 
matching constants are 1 up to NLO in perturbation theory.
A detailed analysis of renormalization for the CI operator 
will be presented in \cite{Gattringer:2003zz} and here we only quote the 
numbers we need.

The renormalization constants were evaluated for the same set of quark masses 
that was also used in the calculation of the matrix elements. When plotted as 
a function of the quark mass, the data
are found to follow a straight line very well and the chiral extrapolation
is straightforward. This procedure was repeated for several values of the
4-momentum $p$. Following \cite{Becirevic:2003pn} we base our calculation on
the numbers extracted at the cutoff, i.e.~at $p^2 = a^{-2}$. 
To get the numbers at 
exactly $a^{-2}$ we linearly interpolated the chirally extrapolated 
values of $Z_V,Z_T$ between the two momenta 
with $p^2$ just above and below $a^{-2}$. The resulting 
numbers for $Z_T$ were then evolved to our target scale of $\mu$ = 2 GeV using 
the renormalization group equation. $Z_V$ remains constant as a 
function of $\mu$ (no anomalous dimension) and we used the value at 
$\mu^2 = a^{-2}$. Our numbers for $Z_V$ and $Z_T$ are listed in Table 
\ref{renormtable}.

\begin{table}
\begin{ruledtabular}
\begin{tabular}{ccccc}
$\beta$ & $Z_V(a^{-1})$ & $Z_T(a^{-1})$ & $Z_T$(2 GeV) &
$Z_T/Z_V$(2GeV)\\ \hline
7.90 & 0.9346(7)  & 1.0598(10) & 1.0247(10) & 1.0964(19) \\
8.35 & 0.9780(14) & 1.0542(19) & 1.0532(20) & 1.0768(37) \\
\end{tabular}
\end{ruledtabular}
\caption{
\label{renormtable}
Results for the renormalization constants. The numbers in parenthesis 
give the scale $\mu$. It is either the cutoff ($\mu = a^{-1}$) or 
$\mu =$ 2 GeV.}
\end{table}

\section{Extraction of the raw data}

If we only consider the vector-meson contribution to the vector-vector (VV) 
and vector-tensor (VT) correlators and contract two of the Lorentz indices, 
we arrive at the following expressions: 
\BEA
g_{\mu\nu} \, \LL 0 \left| \bar q \gamma^\mu q \right| V(p;\lambda) \RR 
\LL V(p;\lambda) \left| \bar q \gamma^\nu q \right| 0 \RR = \EL 
m_V^2 \, f_V^2 \, \LP e_\lambda^\mu e_\lambda^{\nu *} \, g_{\mu\nu} \RP \; ,
\label{VVeq2}
\EEA
\BEA
g_{\mu\nu} \, \LL 0 \left| \bar q \gamma^\mu q \right| V(p;\lambda) \RR 
\LL V(p;\lambda) \left| \bar q \sigma^{\alpha\nu} q \right| 0 \RR = \EL 
i \, m_V^{} \, f_V^{} \, f_V^\perp(\mu) \, p^\alpha \,
\LP e_\lambda^\mu e_\lambda^{\nu *} \, g_{\mu\nu} \RP \; ,
\label{VTeq2}
\EEA
where $e_\lambda^\mu e_\lambda^{\nu *} g_{\mu\nu} = -3$. 
Taking the ratio of these (vector-tensor/vector-vector), we have 
\BE
R^\alpha = \frac{i p^\alpha f_V^\perp(\mu)}{m_V^{} f_V^{}} \; ,
\EE
with one remaining Lorentz index from the tensor current. It is this ratio 
we wish to extract in our lattice calculation. 

For the non-zero-momentum correlators, however, there is an additional 
complication, which arises because the quark sources are smeared in the 
spatial directions. Since the smearing is not performed in the time direction 
as well, Lorentz symmetry is lost, affecting the $e_\lambda^\nu p_\nu = 0$ 
condition (checking this relation explicitly with the different Lorentz 
components of the $\vec p \not= 0$, VT correlators, we find the discrepancy to 
be small, but significant). More explicitly, we have for the VT correlator 
\BEA
\LL 0 \left| (U \bar q) \gamma^\mu (U q) \right| V(p;\lambda) \RR 
\LL V(p;\lambda) \left| \bar q \sigma^{\alpha\nu} q \right| 0 \RR = \EL 
i \, m_V^{} \, f_V^{} \, f_V^\perp(\mu) \, Z_U(p) \,
e_\lambda^\mu \LP e_\lambda^{\nu *} p^\alpha - 
e_\lambda^{\alpha *} p^\nu \RP \; ,
\label{VTeq3}
\EEA
where $U$ represents the smearing operator and $Z_U(p)$ is the 
corresponding factor for the amplitude. For the $\alpha=j$ case, we may no 
longer sum over all of the Lorentz indices, $\mu=\nu$, and still expect the 
second term to vanish. We may, however, consider only the contributions where 
$\mu = \nu = k \not= j$ and $p^k=0$. Then, using only the $\nu = k$, $p^k=0$ 
contributions for the VV correlator as well, we have the appropriate 
cancellations in the ratio $R^j(\vec p)$. This finite-momentum ratio 
yields much larger statistical errors than the zero-momentum ratio, 
$R^0(0)$, and we pursue this approach only as a consistency check. 

We note here that the smearing has no effect upon the zero-momentum results 
since these have contributions only from the spatial components of the 
vector-current source and the smearing amplitude, $Z_U(p)$, is the 
same for all spatial directions at $\vec p = 0$. 

The Euclidean lattice formulation provides a natural selection of the 
desired vector-meson contribution to our correlators; we simply need to 
ensure that our two currents are separated by a large enough distance in the 
imaginary time direction. 
The vector-vector correlator, which corresponds to Eq.\ (\ref{VVeq2}), thus 
reads 
\BE
C^{VV} (\vec p\,;t) \; = \; 
\sum_{\vec x}  \; e^{i\vec p \cdot\vec x}  \;
\LL \LP \bar\psi \gamma_k \psi \RP_{0,0}
\LP \bar\psi \gamma_k \psi \RP_{\vec x,t} \RR \; . 
\label{vector}
\EE
We project to definite momentum $\vec p$, using the necessary phase factor, 
and sum over the relevant, spatial Lorentz indices, $k$ ($p_k = 0$). 
For the tensor current we have to distinguish two different cases:
When the open tensor index is time ($=4$) we consider the temporal correlator
$C^{VT}_4$ defined as ($k$ is summed)
\BE
C^{VT}_4 (\vec p\,;t) \; = \;
i \sum_{\vec x} \; e^{i\vec p \cdot\vec x} \; 
\LL \LP \bar\psi \gamma_k \psi \RP_{0,0}
\LP \bar\psi \sigma_{k 4} \psi \RP_{\vec x,t} \RR \; .
\label{tensor4}
\EE
For open spatial tensor indices $j = 1,2,3$ we study the
spatial correlator $C^{VT}_j$ defined as
\BE
C^{VT}_j (\vec p\,;t) \; = \;
\sum_{\vec x} \; e^{i\vec p \cdot\vec x} \; 
\LL \LP \bar\psi \gamma_k \psi \RP_{0,0}
\LP \bar\psi \sigma_{k j} \psi \RP_{\vec x,t} \RR \; ,
\label{tensorj}
\EE
where $p_j \not= 0$ and the sum over $k$ is restricted to the two values 
different from $j$, such that $p_k = 0$. 
The temporal and spatial correlators $C^{VT}_4$ and $C^{VT}_j$ differ in 
two respects: Firstly, the spatial correlator is non-vanishing
only for non-zero momenta, while 
the temporal correlator gives a contribution also at zero momentum.
Secondly, the spatial and temporal correlators differ by their 
time reversal symmetries. In particular one finds:
\BEA
C^{VV} (\vec p\,;t) & = &  A^{VV}(\vec p\,) 
\LP e^{-E_V^{}(\vec p\,)t} + e^{E_V^{}(\vec p\,)(t - T)} \RP + ... \; ,
\nonumber
\\
C^{VT}_4 (\vec p\,;t) & = &  A^{VT}_4(\vec p\,) 
\LP e^{-E_V^{}(\vec p\,)t} - e^{E_V^{}(\vec p\,)(t - T)} \RP + ... \; ,
\nonumber
\\
C^{VT}_j (\vec p\,;t) & = &  A^{VT}_j(\vec p\,) 
\LP e^{-E_V^{}(\vec p\,)t} + e^{E_V^{}(\vec p\,)(t - T)} \RP + ... \; .
\nonumber
\\
\label{larget}
\EEA
The dots indicate corrections due to excited 
states which play a role only at small time separations 
but become suppressed exponentially for larger values of $t$. 

Let us have a first look at our data for $C^{VV}$ and $C^{VT}$, separately. 
This serves to check how well the expected behavior 
of Eqs.~(\ref{larget}) is seen and how serious the
effects of excited states are. For this test and also for the extraction of 
the data we fold all correlators about $T/2$ according to their time-reversal 
symmetries and average them appropriately. In the plots we thus show only the 
folded and averaged correlator in the range $t = 0 \; ... \; T/2$. 

\begin{figure}
\resizebox{3.2in}{!}{\includegraphics{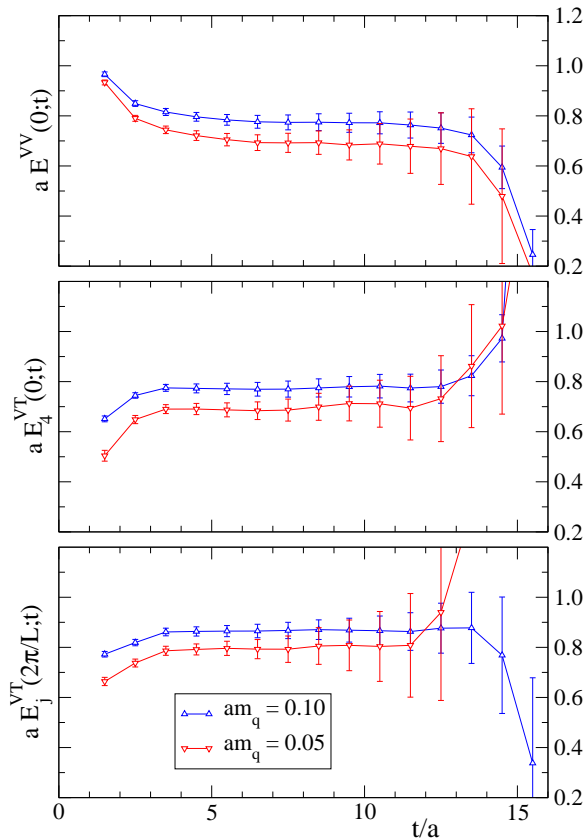}}
\caption{
\label{effmass}
Effective masses for the vector-vector (top plot), 
temporal vector-tensor (middle plot) and
spatial vector-tensor (bottom plot) correlators. 
The effective mass for the spatial vector-tensor correlator is shown
for momentum $\vec p\,^2 = (2\pi)^2/L^2$, while the other two 
effective masses are for zero momentum. The data shown here are for the
$16^3\times32$ lattices with $a = 0.15$ fm and two values of the bare 
quark mass, $am_q = 0.05,0.10.$ }
\end{figure}

In Fig.~\ref{effmass} we show effective mass plots,
i.e.\ $\ln[ C(t)/C(t+1)]$, for our folded correlators defined in 
Eqs.~(\ref{vector}) -- (\ref{tensorj}). The effective mass $E^{VV}$
for the vector-vector correlator 
(top plot) and the effective mass $E^{VT}_4$ for the  
temporal vector-tensor correlator (middle plot) are shown for
zero momentum, while the effective mass $E^{VT}_j$ for the
spatial vector-tensor correlator (bottom plot)
is shown for momentum $\vec p\,^2 = (2\pi)^2/L^2$. The data for the 
figure are from the
$16^3\times32$ lattices with $a = 0.15$ fm and two values of the bare 
quark mass, $am_q = 0.05,0.10$. The spatial correlator has been 
averaged over all three components $j = 1,2,3$. However, the scatter 
of the results for the individual components $j$ is small. 

Both vector-tensor correlators show plateaus between
$t/a = 4$ and $t/a = 12$ and the vector-vector correlator for 
$t/a = 6 \; ... \; 12$. For smaller time separations it is obvious
that contributions from excited states are important. 
For $t/a$ larger 
than 12 we observe two effects: Firstly, an increase of the error
bars which is simply an effect of the reduced signal to noise ratio 
for longer propagation in $t$ and, secondly, a systematic deviation from 
the plateau due to the sinh / cosh behavior of the correlators. 
For all quark masses we looked at we were able to identify stable plateaus.
When comparing our different results for the vector meson energy
$E_V(\vec p\,)$ which determines the time behavior of all three 
correlators [see Eq.\ (\ref{larget})] we found that the values of 
$E_V(\vec p\,)$ agreed well for the different 
operators. Note that the spatial vector-tensor correlator shown in 
the bottom plot of Fig.\ \ref{effmass} is for nonzero momentum and 
thus its plateau is slightly higher than for the other two correlators. 

Once the amplitudes are known we can form the temporal and spatial 
ratios $R_4$ and $R_j$ defined as
\BEA
R_4(\vec p\,) & = & A^{VT}_4(\vec p\,) / A^{VV}_{}(\vec p\,) \; ,
\nonumber
\\
R_j(\vec p\,) & = & A^{VT}_j(\vec p\,) / A^{VV}_{}(\vec p\,) \; .
\EEA
We have already remarked that $R_4$ is non-vanishing also at zero momentum, 
while $R_j$ gives a contribution only for non-zero momenta. Since the quality 
of lattice correlators generally is decreasing with increasing momenta we 
evaluate $R_4$ at zero momentum and $R_j$ at the smallest possible momentum, 
i.e., $\vec p\,^2 = (2\pi)^2/L^2$. The ratios $f_V^\perp/f_V$ are then 
obtained from
\BE
R_4(0) = \frac{f_V^\perp}{f_V^{}} \; \; , \; \; 
R_j(\vec p\,) = \frac{p_j \; f_V^\perp}{m_V^{} \; f_V^{}} \; .
\EE
For the vector mass $m_V$ in the second equation we may use the 
value obtained from our fits to the same correlators; i.e., 
$m_V^{} = \sqrt{E_V^{\,2}(\vec p\,)-\vec p^{\,2}}$. 
Note that here $f_V^{}$ and $f_V^\perp$ are bare lattice quantities which 
need to be renormalized.

\section{Results}

In order to extract the appropriate values for the amplitudes, $A^{VV}_{}$ 
and $A^{VT}_{\nu}$, from the two separate correlators, we construct a 
$\chi^2$ measure of the form
\BEA
\chi^2 &=& \sum_{I,J} \sum_{t^I,t^J} 
\LS C^I(\vec p\,;t^I) - C_{exp}^I(\vec p\,;t^I) \RS \EL 
&& \; \times \; S_{I,J}^{-1}(t^I,t^J) 
\LS C^J(\vec p\,;t^J) - C_{exp}^J(\vec p\,;t^J) \RS \EL 
&=& F \LP A^{VV}_{}(\vec p\,), A^{VT}_\alpha(\vec p\,), 
E_V^{}(\vec p\,) \RP \; , 
\EEA
where the ``expected'' values, $C_{exp}^I(\vec p\,;t)$, are given by the 
appropriate expressions in Eqs.~(\ref{larget}) and $S^{-1}_{}$ is the inverse 
of the covariance matrix. 
We extract the three parameters by minimizing the 
above function, taking care to use large enough minimum $t^I$ values, 
thereby avoiding excited-state contributions. See Table \ref{fits_table} 
for the results of the zero-momentum fits.

Since we are interested in the ratios of the resulting amplitudes, we 
repeat these fits within a single-elimination jackknife routine, 
providing the errors for $f_V^\perp/f_V^{}$ (see the final column of 
Table \ref{fits_table}). 

\begin{table}
\begin{ruledtabular}
\begin{tabular}{cccccc}
$a$ (fm) & $am_q$ & $t^{VV}$ & $t^{VT}$ & $\chi^2$/d.o.f. & $R_4(0)$ \\ \hline
0.15 & 0.02 & $8-12$ & $4-12$ & 10/11 & 0.678(36) \\
$''$ & 0.03 & $8-13$ & $4-13$ & 11/13 & 0.680(24) \\
$''$ & 0.04 & $8-15$ & $4-14$ & 14/16 & 0.688(18) \\
$''$ & 0.05 & $8-16$ & $4-15$ & 18/18 & 0.696(14) \\
$''$ & 0.06 & $8-16$ & $4-15$ & 18/18 & 0.707(10) \\
$''$ & 0.08 & $8-16$ & $4-15$ & 17/18 & 0.725(7) \\
$''$ & 0.10 & $8-16$ & $4-15$ & 16/18 & 0.740(5) \\ \hline
0.10 & 0.04 & $9-16$ & $6-15$ & 21/15 & 0.716(9) \\
$''$ & 0.05 & $9-16$ & $6-15$ & 23/15 & 0.722(7) \\
$''$ & 0.06 & $9-16$ & $6-15$ & 25/15 & 0.729(6) \\
$''$ & 0.08 & $9-16$ & $6-15$ & 29/15 & 0.744(4) \\
\end{tabular}
\end{ruledtabular}
\caption{
\label{fits_table}
Selected results (those used for the quark mass extrapolation) of the 
three-parameter fits for the zero-momentum correlators. 
}
\end{table}

\begin{figure}[t]
\resizebox{3.2in}{!}{\includegraphics{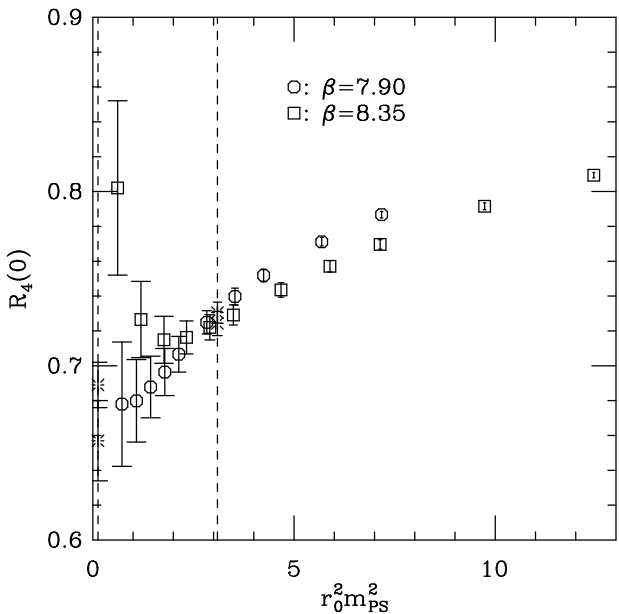}}
\caption{
\label{amp_ratio_vs_mps2}
Ratio of vector meson couplings (at the lattice scale,
$\mu=1/a$) as a function of $r_0^2 m_{PS}^2$ for the
two lattice spacings. The dashed vertical lines denote the $r_0^2 m_\pi^2$
and $r_0^2 m_{s\bar s}^2$ ($m_{s\bar s}^2 = 2m_K^2 - m_\pi^2$) physical
values. The bursts denote the mass interpolations at $m_{s\bar s}^2$ and
extrapolations to $m_\pi^2$.
}
\end{figure}

Figure \ref{amp_ratio_vs_mps2} displays the ratios of the vector meson
couplings as a function of the dimensionless quantity, $r_0^2 m_{PS}^2$.
For a wide range of quark masses the data display only a slight curvature 
and are quite compatible with a linear behavior. 
Only for the smallest quark masses do the data move upwards. 
Simulations on smaller physical volumes ($8^3\times 24$ and $12^3\times 24$) 
at a fixed lattice spacing ($a$ = 0.15 fm) 
suggest that the upward trend of the ratio at small quark 
masses is a finite-size effect. This interpretation is supported 
by the observation that also in spectroscopy calculations of nucleons and 
their excitations the finite size effects set in at the same values of the 
quark mass \cite{Gattringer:2002xg,Gattringer:2002sb,Gattringer:2003bp}. 
In order to avoid including such effects, we perform the (fully correlated) 
extrapolation to light quark mass without some of the lightest bare quark 
masses. 
\begin{table}[b!]
\begin{ruledtabular}
\begin{tabular}{cccc}
 & \vline & $a=0.15$ fm & $a=0.10$ fm \\ \hline
$\LP\frac{f_\phi^\perp}{f_\phi^{}}\RP_{\overline{\rm MS}}$(2 GeV) & \vline & 0.801(7) & 0.780(8) \\
 & \vline & & \\
$\LP\frac{f_\rho^\perp}{f_\rho^{}}\RP_{\overline{\rm MS}}$(2 GeV) & \vline & 0.720(25) & 0.742(14) \\
\end{tabular}
\end{ruledtabular}
\caption{
\label{msbar_ratios}
$f_V^\perp / f_V^{}$ values in the $\overline{\rm MS}$ scheme at $\mu=2$ GeV 
as determined from the lattice ratio $R_4(0)$.
}
\end{table}
This is necessary only for our fine lattices ($\beta=8.35$), where the three 
smallest quark masses give rise to pseudoscalars with $m_{PS} L < 4$. 
As can be seen in the plot, it is exactly these three data points which show 
finite-size effects and we therefore exclude them from the chiral 
extrapolation. For our larger lattices ($\beta=7.90$), we always have 
$m_{PS} L > 4$; no finite-size effects are visible and all data points are 
included in the fit. 
On the heavy end, all masses with $r_0^2 m_{PS}^2 > 5$ are also excluded 
from the extrapolation. 
To obtain the data at the strange quark mass we interpolate the neighboring 
data linearly. 

In Table \ref{msbar_ratios} we present our final results for the ratios of 
the couplings, matched to the $\overline{\rm MS}$ scheme and evolved to the 
scale of 2 GeV. These results are obtained from the zero-momentum correlator 
ratios, $R_4(0) \times Z_T/Z_V (2 \, {\rm GeV})$.

\begin{figure}[t!]
\resizebox{3.2in}{!}{\includegraphics{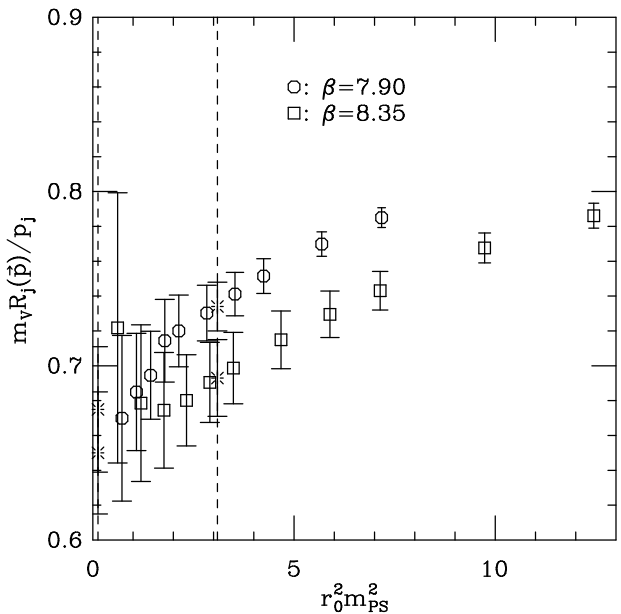}}
\caption{
\label{amp_ratio_i_vs_mps2}
Similar plot to that of Fig.~\ref{amp_ratio_vs_mps2}, this time for the 
finite-momentum ratio $m_V^{} R_j(\vec p\,)/p_j$, where 
$|\vec p\,| = p_j = 2\pi/L$. The symbols have the same meaning as before. 
}
\end{figure}

Since we include no improvement in our current operators, we may expect a 
lattice spacing dependence of ${\cal O}(a)$. However, since we only have 
results for two values of the lattice spacing, a continuum extrapolation 
becomes problematic.
\begin{table}[b!]
\begin{ruledtabular}
\begin{tabular}{cccc}
 & \vline & $a=0.15$ fm & $a=0.10$ fm \\ \hline
$\LP\frac{f_\phi^\perp}{f_\phi^{}}\RP_{\overline{\rm MS}}$(2 GeV) & \vline & 0.805(15) & 0.746(24) \\
 & \vline & & \\
$\LP\frac{f_\rho^\perp}{f_\rho^{}}\RP_{\overline{\rm MS}}$(2 GeV) & \vline & 0.740(39) & 0.700(38) \\
\end{tabular}
\end{ruledtabular}
\caption{
\label{msbar_ratios_ij}
$f_V^\perp / f_V^{}$ values in the $\overline{\rm MS}$ scheme at $\mu=2$ GeV 
as determined from the lattice ratio $R_j(2\pi/L)$. 
}
\end{table} 
We do note that the two results for the $\rho$ 
meson are consistent with each other and they are also in agreement with the 
results of other lattice calculations 
\cite{Capitani:1999zd,Becirevic:2003pn}: 
$0.72(2)(^{+2}_{-0})$ in Ref.~\cite{Becirevic:2003pn}. 
The trend of the $\phi$ results suggests a continuum value below our 
$\beta=8.35$ result of 0.78, in rough agreement with the 0.76(1) result 
of Ref.~\cite{Becirevic:2003pn}.

If we normalize using the experimental value of 
$f_{\rho^\pm}^{\,\rm exp.} \approx 208$ MeV 
\cite{Hagiwara:2002pw,Becirevic:2003pn}, 
we find good agreement with the QCD sum rule result
\cite{Ball:1996tb,Ball:1998kk,Bakulev:1999gf}: 
$f_\rho^\perp/f_{\rho^\pm}^{\,\rm exp.} = 0.74(3)$. 
Such comparisons, however, are difficult to assess due to the different 
systematics in the two approaches (e.g., our lattice calculation is 
quenched).

The results for the finite-momentum correlator ratio, $R_j(\vec p\,)$, are 
shown in Fig.~\ref{amp_ratio_i_vs_mps2}. The quark-mass interpolations and 
extrapolations are performed just as before for the zero-momentum ratios. 

Table \ref{msbar_ratios_ij} displays the renormalized coupling ratios 
obtained from the non-zero-momentum correlator ratios. The errors are 
significantly larger for these than those for the zero-momentum results. 
However, the results of this consistency check are compatible with 
those from the zero-momentum ratios ($< 1.5\sigma$ even for the jackknifed 
difference at and below the strange-quark mass). 

Let us briefly summarize our findings. We calculate $f_V^\perp / f_V^{}$ 
going considerably closer to the chiral limit 
(at least on our larger lattices where the results do not display 
significant finite-volume effects) 
than previous calculations. 
This allows us to monitor the mass dependence of this ratio of couplings 
over a much larger mass range. We find the quark mass dependence to be 
relatively weak, as in QCD sum rule calculations. 
The values we obtain agree well with the extrapolated values of other 
lattice calculations at larger quark masses. 
We include a consistency check using spatial tensor correlators at finite 
momentum and find compatible results. 
Our final numbers are 
$f_\rho^\perp / f_\rho^{} = 0.720(25), \, 0.742(14)$ and 
$f_\phi^\perp / f_\phi^{} = 0.801(7), \, 0.780(8)$ at 
$a = 0.15, \, 0.10$ fm, respectively.

\vspace{-0.5cm}
\begin{acknowledgments}
\vspace{-0.4cm}
We would like to thank 
Damir Becirevic, 
Peter Hasenfratz, 
Philipp Huber, 
Christian Lang, and 
Ferenc Niedermayer 
for helpful discussions. 
The calculations were performed on the Hitachi SR8000 at the Leibniz 
Rechenzentrum in Munich and we thank the LRZ staff for training and support. 
This work was supported by the DFG-Forschergruppe 
Gitter-Hadronen-Ph\"anomenologie. 
C.~Gattringer acknowledges support by the Austrian Academy of Sciences 
(APART 654). 
\end{acknowledgments}

\bibliography{vector_tensor}

\end{document}